\RequirePackage{silence} 
 \documentclass[11pt,twoside]{article} \newcommand\aut {Leonid A.~Levin}
 \newcommand\ttl {Assumptions of Randomness in Cosmology Models}
 \usepackage[margin=1in]{geometry} \usepackage[unicode,pdftex]{hyperref}
 \usepackage{mmap} 
\begin{document} \frenchspacing
 \newcommand\hreff[1]{\href {https://#1} {https://#1}} \newcommand\w\omega 
 \renewcommand\smile{\mbox{:-}}

 \title{\vspace*{-76pt}\ttl}\author{\aut~ (\hreff{www.cs.bu.edu/fac/Lnd/})}
 \date{Boston University} 
 \maketitle\vspace{-25pt}

\begin{abstract} Non-compact symmetries cannot be fully broken by randomness
 since non-compact\\ groups have no invariant probability distributions.
 In particular, this makes trickier\\ the ``Copernican'' random
 choice of the place of the observer in infinite cosmology models.

 {This problem may be circumvented with what topologists call
  {\em pointed spaces}.\\ Then randomness will be used only
  in building (infinite) models around the\\ pre-designated
 ``observation point'', that thus would not need to be randomly chosen.}

 Additional complications come from the original randomness
 possibly being hidden. P. Gacs and A. Kucera proved that every
 sequence can be algorithmically generated from a random one.
 But Vladimir V'yugin discovered that randomized algorithms can with
 positive probability generate uncomputable sequences that are not
 algorithmically equivalent to any random ones. \end{abstract}

Physics laws allow all sorts of events and histories. In infinite cosmology
models each one of them rolls out infinitely often: weirdest miracles are daily
routine in some places. To understand our boring world one must assume our
place is rather generic. So, theories assume some events happen at random,
others are derived from them by physical equations or other
algorithms. \mbox {In fact,} \cite {gacs,Kr} prove that any sequence is
 computable from an algorithmically random one.\footnote
 {Algorithmic randomness (see: \cite{K,KU}) of a digital sequence $\w$ does not
  quite assure it obeys all probability $1$ math laws (only computable ones).
  Yet, it assures (see: \cite {L84}) that $\w$ either obeys all such laws
  or has an infinite information about some mathematically
  definable object. The latter case would be really weird. There are
  no ways for such $\w$ to be generated; see this informational version
  of Church-Turing Thesis discussed at the end of \cite {L13}.}
 However, the theories of this type can only be complete if they specify
which exactly are the random events and what is their probability
distribution. These events need not be directly observable, rather just
be a part of the model from which the observations are derived.
 By \cite{vv},~resulting observables, unlike random ones, could
 be very diverse in algorithmically invariant properties.

But the specification of this set of ``external'' events (not meant to be
derived in the model, but rather assumed random) is often glossed over.
 Someone like Laplace, or another classical deist, might assume this to
be the set of positions and speeds of all particles at the initial state
of the Universe. The Creator would set them up, then retire, leaving the
rest to the equations of physics. But this choice is quite arbitrary. The
Creator could as easily set up the final state of the Universe at random.
(As all equations known to Laplace were time-reversible.) Then the
initial state would be very non-random! This may just flip the timeline,
but even more elegant would be to set externally both initial {\em and
final} positions of particles (and those on borders, if any), but no
speeds. After all, the positions of a thrown stone at two moments of
time determine its trajectory in-between. Different choices of events
designated as external and random, lead to quite different predictions.

The model also needs to describe what is a legitimate set of
observations, which includes specifying the observer (say, the humanity
with its followers). Some models have the Universe as a whole in a pure
state, any perceived entropy is due to the whole being unobservable: each
specific observer can see only a part,\footnote
 {The unitary evolution of the whole Universe preserves its pure state.
Observers can be separated by horizons but even if not, the heat escaping
to the horizon leaves the remaining observable subsystem in a mixed high
entropy state. If the Universe stops expanding and recollapses,
the escaped heat returns, and being entangled with the left behind
matter, might restore its pure state. However, imagining the mechanism
for this is tricky, which provokes switching to more complicated infinite
models that face problems, including those discussed below.}
 which is in a mixed state. (In QM, unlike Classical Physics, parts can
have higher entropy than the whole.) Then the choice of the observer is
actually the source of the randomness. While the theory may be largely
indifferent to that choice, it may be impossible to formulate without one:
remote observers cannot communicate to combine experiences. The choice of
the observer is as much a part of a complete model, as is the state of
the rest of the universe.

\vspace{6pt} {\bf A technical difficulty} comes up here. While the
observable part of the world is finite, many models include it in a
greater infinite realm, beyond our reach. How to choose our (observer's)
place in this realm at random is puzzling. Say, in a most primitive
setting, how to choose a random point in a pattern on an infinite
Euclidean space? It carries no uniform probability distribution. (In
other models, where ours is one of the infinitely many ``bubbles,'' it
may be even much trickier.) A way around this technicality can lie in the
reverse order of choices. Instead of first building an infinite universe
and then choosing the place of the observer there, we can first designate
the observer, then build an infinite universe around her.\footnote
 {One may argue that computing probability of any {\em finite} event requires
  only placing the observer in a finite region, which can be done at random.
  However this does not provide a complete model of the infinite universe.
  Some global properties do not manifest in any finite region.
  And a compact event $X$ is the intersection of a monotone family of
  clopen (finite) events $X_i$. Its probability is $\inf_i$Prob$(X_i)$.
  But $X_i$s must all agree on the observer's surroundings.
  So that would anyway amount to setting the observer first
  and then using randomness to generate the universe around.}
 This circumvents the lack of uniform distributions in infinite domains
(and tickles our ego, besides \smile). Such structures are called {\em
pointed spaces} in topology. Among the challenges is the need to expand
the universe not only to the observer's present and all future times, but
also to her whole past (if there are any causal relations with the past).

This brings up a question, what is an observer? Its many meanings depend
on what symmetries are meant to be broken by choosing this sort of a
``gauge''. In quantum physics observer can mean a macroscopic system
brought into an entanglement with the observed microscopic variables.
``Macroscopic'' is a vague term with unclear relevance. Sometimes it is put
in relation with the Planck mass (e.g., by R.Penrose). But it is unclear
why a microscopic bacterium, much lighter than the Planck mass, cannot
perfectly serve as an observer. (It could transmit its observed data to
its descendants, which may include members of Royal Society.~\smile)

What seems relevant here, is that the observed data, unlike generic
quantum states, can be reliably copied, preserved, transmitted, etc. For
this an observer needs some sophistication, access to mechanisms for
error-correction, self-preservation, etc.
 (This requires free energy flow: so gravity, its ultimate source, may
play a role for observers, even if not via Planck~mass. To enter a mixed
state, observers need to discard parts of the system. Thus they consume
free energy and expel heat to some horizon: cosmological or black hole,
or some such. This generates entropy, both of the observer and of the
horizon if taken separately, though entropy formulas for those horizons
are often used somewhat confusingly.\footnote
 {When a neutron star collapses into a black hole, its mass changes
little, radius shrinks twice or so, but its Bekenstein-Hawking entropy
upper bound (= area) grows by 20 orders of magnitude, approaching the
combined entropy of all neutron stars in the universe. This sounds strange.
 The collapse is unitary, so black hole v.Neumann entropy is bounded by
the entropies of the collapsed star and emitted entangled radiation.

This is $O(m^{1.5})$ for black holes of the irreducible mass $m$
(= radius/2), and is observable, not speculative.\\
 (Ordinary matter cannot have entropy $r^{1.5}$ within radius $o(r)$
in Planck units: by Pauli exclusion for fermions,\\ Stefan-Boltzmann
energy density for photons, Schwarzschild radius limit, etc.)

The $4\pi\;m^2$ bound is inspired by the GR area monotonicity theorem
(holds even locally). I do not know if GR in asymptotically flat vacuum
has global counterexamples to a stronger monotonicity with $m^{1.5}$
replacing $m^2$.\\
 The coarse-grained bound $m^2$ assumes the actual state is randomly
chosen among all possible states consistent with the remaining observables.
This is plainly false, though the state is speculated to become random
in $10^{67}$ years by radiating $m^2$ Hawking quanta entangled with it.
(Assuming black hole radiates, has temperature, and it is $1/m$.) })
  Such features are readily present in life. Our world is in a state that
harbors life, however mysterious is life's origin. The present living
creatures developed by evolution. But to start, the evolution needs
systems capable of copying themselves, along with~accumulated random
mutations. The first such systems could not be produced by evolution, so
must appear spontaneously. The minimal complexity $C$ of such systems may
be significant, and the chance $P$ of their spontaneous generation too
tiny: $1/P$, exponential in $C$, exceeding cosmological scales.

\vspace{1pc} If so, how to deal with these tiny probabilities?
It may be that the laws of physics are fine-tuned to boost them.
 But it may be that this fine-tuning requires high complexity of such
fine-tuned laws, which just transfers the tiny probability issue from
live systems to laws of physics.

(I avoid the popular but unclear term ``Anthropic principle''. As was
said ``To discover a land means to look upon it with the eyes of a
European, preferably British.'' Is fluent English required for being
fully Anthropic? (\smile) I assume for observers only the seemingly
relevant abovementioned ``copyright'' for observed data. I even ignore
another frequent vague assumption -- the free will.\footnote
 {Stretching the relevance, let me at least explain how I understand this
``free will'' notion.\\
 Physics equations generate the world from some random variables.
Observations change their original \mbox{probability} distribution, conditioning
it on the observed knowledge. (Observations reveal only some complicated
derivative \mbox{restrictions} on the original, not directly observed, random
variables. This makes tricky guessing our environment in~some useful
ways: we use ``Occam Razor,'' analogies, etc., instead of computing
probabilities directly.)\\ We model some systems describing their behavior
or a probability distribution under which that behavior is random.

An impossibility of having an adequate model is expressed as system's free will.
A modeling entity needs to exceed the modeled one in complexity,
so cannot model itself, must accept its own free will. And free will is
``contagious'': extends to anything interacting with a free-willed~entity.
 As an example of free will subtlety let me bring Hillel's Golden Rule (expect
yourself to be treated the way you treat others). It can be promoted by
effective systems of justice (expensive and tricky to design) that reveal,
judge, and reciprocate what we do to each other. It can also be promoted by
a PR-induced faith that such a system exists, working in some mysterious way.
But PR and mysteries, overused for corrupt purposes, meet widespread skepticism
nowadays which impairs that approach.

The meaning of free will can clarify another approach:
{Observing my own \bf unpredictable} actions, I learn more about myself.
I interact with fellow humans who share with me origins, environment, and many
qualities.\\ Thus learning about myself, I indirectly learn about them, too.
This knowledge changes the probability distribution on my environment. For
instance, my kind or wicked acts, besides their immediate effects, reveal to me
more aspects of human nature, of which mine is the most intimately visible
example. This subtle ``two-way street'' effect of our free will (on human
nature we get to expect) sheds some light on the faith in the Golden Rule.})

However, the mystery seems to soften with the abovementioned priority of
choosing the observer. If models start with designating an observer (and
then choosing her details and building a universe around) then the worlds
with no observers are excluded before computing the probabilities.

Such settings may also help with another issue. Some fundamental laws of
physics seem to yield paradoxes or even conflict with each other. But it may
suffice for our needs if the laws are only {\em approximately} sound. For
instance, the observer cannot be absolutely reliable. It may spontaneously
tunnel into something entirely different, albeit with an exponentially small
probability.\\ A theory may be clear if such effects are ignored, assuming the
observer behaves ``as advertised''. And it may be O.K. for the theory to become
incomprehensible if negligible likelihoods, such as, e.g., observer's drastic
tunneling, must be accounted for.

\vspace{-7pt}\paragraph {Acknowledgments:} Thanks to Charles Bennett and
Alexander Vilenkin for insightful criticism.

\end{document}